\documentclass[11pt]{article}
\usepackage[utf8]{inputenc}
\usepackage[T1]{fontenc}
\usepackage{lmodern}

\usepackage{graphicx}
\usepackage{amsmath}
\usepackage{amssymb}
\usepackage{booktabs}
\usepackage{caption}
\usepackage{geometry}
\usepackage{subcaption}
\usepackage{longtable}

\usepackage[hyphens]{xurl}
\usepackage{hyperref}
\usepackage{setspace}

\usepackage[backend=biber, style=numeric, sorting=none]{biblatex}
\hypersetup{
    colorlinks=true,
    linkcolor=blue,
    citecolor=green,
    urlcolor=magenta,
    pdftitle={Displacement Extraction from Myanmar Earthquake Footage Video},
    pdfauthor={Fuhua Zheng, Jianhao Gao, Haoran Meng, Chaofeng Wang},
    pdfkeywords={Earthquake, Computer Vision, Ground Motion},
    bookmarksopen=true,
    bookmarksnumbered=true
}

\title{Video-based Direct Time Series Measurement of Along-Strike Slip on the Coseismic Surface Rupture During the 2025 Mw7.7 Myanmar Earthquake}

\author{
  Jianhao Gao\textsuperscript{1,$\dagger$} \and
  Fuhua Zheng\textsuperscript{3,$\dagger$} \and
  Chaofeng Wang\textsuperscript{1,2,$\ast$} \and
  Haoran Meng\textsuperscript{3,$\ast$}
}

\date{May, 2025} 

\begin{document}
\onehalfspacing
\maketitle

\begin{center} 
  \textsuperscript{1}M.E. Rinker, Sr. School of Construction Management, University of Florida, USA \\
  \textsuperscript{2}Department of Civil \& Coastal Engineering, University of Florida, USA \\
  \textsuperscript{3}Department of Earth and Space Sciences, Southern University of Science and Technology, China \\
  \vspace{0.5em} 
  \textsuperscript{$\dagger$}These authors contributed equally to this work. \\
  \textsuperscript{$\ast$}Corresponding authors: \href{mailto:chaofeng.wang@ufl.edu}{\texttt{chaofeng.wang@ufl.edu}}, \href{mailto:menghr@sustech.edu.cn}{\texttt{menghr@sustech.edu.cn}}
\end{center}
\vspace{1em} 

\begin{abstract}
This study presents a time-resolved analysis of coseismic lateral surface rupture along the Sagaing Fault during the Mw 7.7 Mandalay, Myanmar earthquake on March 28, 2025. Leveraging a publicly available Closed-Circuit Television (CCTV) footage alongside on-site measurements, we show the first in-situ high sampling rate direct measurement of a coseismic slip evolution of a fault during an earthquake. Our work comprises four primary stages: data acquisition, video pre-processing, object tracking, and physical displacement estimation. Video pre-processing includes camera stabilization and distortion correction. 
We then track pixel-level movements of selected reference points using two complementary computer-vision approaches -- a traditional grayscale template matching algorithm \cite{opencv_library} and a state-of-the-art vision transformer multi-object tracking algorithm \cite{karaev23cotracker, karaev24cotracker3}, and verify both profiles against meticulous manual frame-by-frame measurements, with results that closely match one another. 

Finally, we translated those pixel displacements into real-world ground movements by calibrating against reference objects whose dimensions were measured on site. Based on the resulting displacement time series, we estimated the critical slip-weakening distance. The resulting high-resolution time series of the along-strike slip, provided in the appendix, offers a critical benchmark for validating dynamic rupture simulations, refining frictional models, and enhancing seismic hazard assessment.

\end{abstract}

\section{Introduction}
Understanding the coseismic rupture process is fundamental to describing fault mechanics, quantifying energy release, and mitigating near‑fault seismic hazards. The slip distribution is one of the key parameters of the rupture inversion. For large earthquakes, kinematic slip distributions are usually inferred from (i) finite‑fault inversion of teleseismic and strong‑motion waveforms \cite{hao2017slip, ye2016rupture, gong2022rupture}; or (ii) joint inversions that combine InSAR, GPS, and field data \cite{tong2010coseismic, yue2022rupture}. Each method faces some limitations due to the data characteristics.

Seismographs usually record the velocity or acceleration of a location, supplying millisecond‑scale temporal resolution, but is an indirect proxy for slip. For a large earthquake with limited seismic stations in the near‑field and most broadband stations hundreds of kilometers from the fault, high‑frequency attenuation is inevitable, causing waveform inversions to be restricted to low‑frequency signals, inhibiting the resolution of details on the fault. Geodetic observations provide spatially dense snapshots of permanent surface deformation, capturing large-scale displacement patterns. However, it typically represents the cumulative displacement before and after the event and do not resolve the temporal evolution of the slip during the rupture process.

At 12:50:54 p.m. on March 28, 2025 Myanmar Standard Time (MST), a magnitude 7.7 strike-slip earthquake struck near Mandalay, Myanmar, causing extensive destruction and human loss across multiple countries. The rupture extended approximately 500~km along the north-south trending Sagaing Fault, a major transform fault cutting through central Myanmar, as confirmed by back projection and satellite observations. The earthquake caused a peak ground acceleration of 0.6231~g and a peak ground velocity of 161.42~cm/s at seismic station NPW, located just 2.75~km from the fault. The finite‑fault inversion result given by USGS \footnote{\url{https://earthquake.usgs.gov/earthquakes/eventpage/us7000pn9s/finite-fault}} reveals peak subsurface slip exceeding 4 m over ~500 km of rupture. Coseismic displacements map along the fault derived from multi-pair feature tracking of Sentinel-2 images \cite{van_wyk_de_vries_2025_15123647} confirmed maximum offsets greater than 4 m and delineated variable slip distribution along the rupture. Ground surface displacement maps near the epicenter in Mandalay show pronounced north–south surface slip \cite{gascoin_2025_15127866}.

The coseismic rupture process of this major earthquake was captured on video at close range. This is possibly the first time such an event has been documented in this manner. A fixed CCTV camera facing directly toward the fault fortuitously recorded the full sequence of ground displacement during this earthquake. This footage reveals how surface co-seismic slip displacement evolved on a sub-second level, which can not be achieved by previous field data and inversion methods.

Taking advantage of this video, we reconstructed the coseismic slip time series using a combination of computer vision techniques and field measurements. The dataset is included in the supplementary tables for potential use in scientific research and engineering applications. The computer vision approach involved camera stabilization and distortion correction, followed by pixel-level (and sub-pixel-level for CoTracker) displacement tracking using different algorithms. The resulting displacement time series was estimated using field-measured reference dimension. The data complements existing geophysical datasets by providing direct measurements for validating and refining kinematic rupture models, particularly in the near field where instrumental coverage is often sparse. It also benefits engineering applications, such as earthquake-resistant design and structural dynamic assessment.

For the remainder of this article, the structure is as follows:
\autoref{sec:methodology} outlines the technical steps involved, including data acquisition, pre-processing, pixel displacement tracking, and post-processing of physical displacement mapping.
This is followed by the results presented in \autoref{sec:result}, which include the tracked pixel displacements and their corresponding real-world estimations. Discussion and conclusion are followed.

\section{Method}\label{sec:methodology}

\subsection{Data Acquisition}

A CCTV camera on a solar photovoltaic farm operated by Green Power Energy Company Limited (Myanmar) fortuitously captured the entire rupture process during the earthquake. The camera faces southward and locates at \(20^\circ 52' 55''\text{N},\ 96^\circ 02' 08''\text{E}\), approximately 123.8~km south of the epicenter and only \(\sim 15\) meters side from the surface fault trace. This footage provides the most detailed visual documentation of an earthquake rupture to date, offering a rare opportunity to observe the rupture process at close range.

The video is available online\footnote{\url{https://www.youtube.com/watch?v=77ubC4bcgRM}} and features high-resolution frames of \(1280 \times 720\) pixels at a frame rate of 30~fps. It records both the initial shaking caused by seismic wave propagation and the subsequent progressive surface dislocation on either side of the fault caused by the rupture.

To extract quantitative information, we conducted field measurements of the physical dimensions of several reference objects visible within the camera's view. These measurements enabled reliable recovery of actual fault-parallel displacements from the 2D video footage. Consequently, we constructed a displacement--time series across the fault, offering new insights into coseismic surface rupture dynamics from a near-field visual perspective.

\begin{figure}[htbp]
\centering
\includegraphics[width=0.55\textwidth]{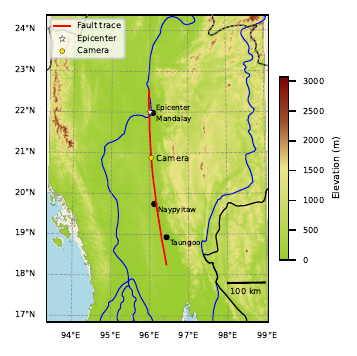}
\caption{
Location of the surveillance camera (yellow circle) and the fault trace (red line) ruptured during the 2025 Mw7.7 Mandalay earthquake. Fault trace is given by back projection. The white star shows the epicenter.
}
\label{fig:location}
\end{figure}

\subsection{Camera Stabilization}

To compensate for inter-frame motion caused by camera shaking, video stabilization was performed using affine transformations estimated from four manually selected reference regions. The first frame in the sequence was chosen as the reference frame.

\begin{figure}[htbp]
\centering
\includegraphics[width=0.9\textwidth]{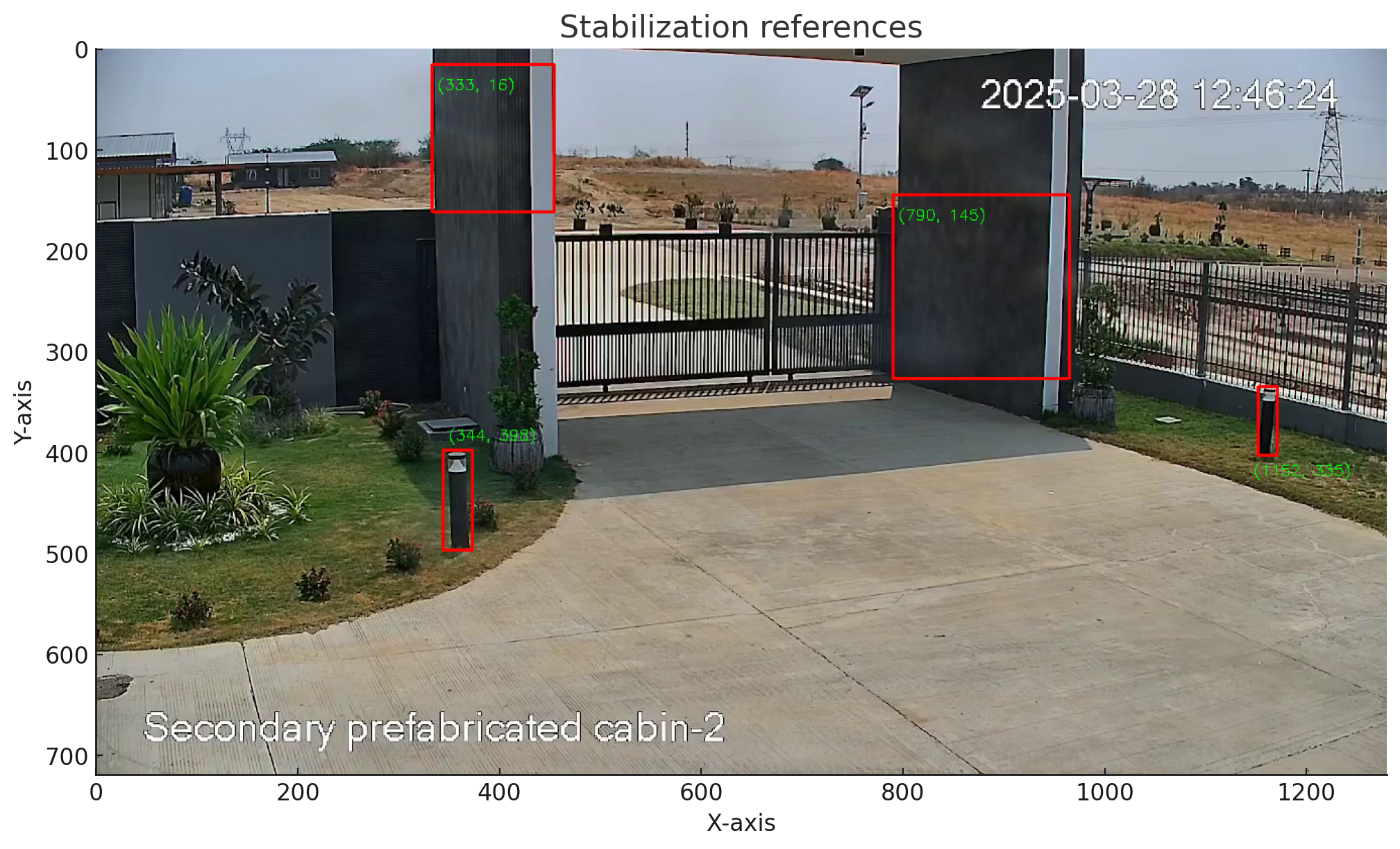}
\caption{Four selected regions were used for video stabilization to correct camera shake. Best-matching areas in each frame were identified via grayscale normalized cross-correlation, followed by affine transformation.}
\label{fig:stabilization}
\end{figure}

As illustrated in \autoref{fig:stabilization}, four rectangular template regions were selected—two on the walls and two on the pathway lights—each containing visually stable and trackable features. For each of these regions, normalized cross-correlation in grayscale was used to locate the best-matching region in every subsequent frame. The center coordinates of the matched regions were extracted and used as control points.

An affine transformation was then estimated by fitting the control points in each frame to their corresponding locations in the reference frame using a least-squares method. This transformation was applied to each frame in the sequence, aligning the four control points to their original coordinates in the first frame, thereby producing stabilized video frames. 

Particularly, we took below pixel regions as reference to stabilize the video: (344, 398, 29, 99), (1152, 335, 19, 68), (333, 16, 121, 146), and (790, 145, 175, 182). Each set of values represents $(x, y, \text{width}, \text{height})$, where $(x, y)$ indicates the coordinates of the top-left corner of the region. These regions are ad hoc selections corresponding to four assumed stable objects: two ground lights and two walls.

\subsection{Distortion Correction}

To correct barrel distortion—visible as curved lines along door edges in \autoref{fig:stabilization} and to align vanishing points for the first and last frames—we tested multiple distortion parameter sets. The final camera model assumes a 90-degree horizontal field of view (FOV) with a principal point at pixel coordinates $(640, 360)$. Distortion correction was applied using OpenCV’s `cv2.undistort', which models radial distortion as:
\begin{align}
x_{\text{distorted}} &= x \left(1 + k_1 r^2 + k_2 r^4 + k_3 r^6\right) \\
y_{\text{distorted}} &= y \left(1 + k_1 r^2 + k_2 r^4 + k_3 r^6\right)
\end{align}
Here, $x$ and $y$ are normalized image coordinates, which are derived by centering the pixel coordinates on the optical axis and scaling by the focal length. The variable $r^2 = x^2 + y^2$ is the squared radial distance from the center, and $k_1, k_2, k_3$ are the radial distortion coefficients. For a more detailed explanation of the distortion model and parameter definitions, please refer to the OpenCV documentation for camera calibration \cite{opencv_library}.

\begin{figure}[htbp]
  \centering
  \begin{subfigure}[b]{0.9\textwidth}
    \centering
    \includegraphics[width=\textwidth]{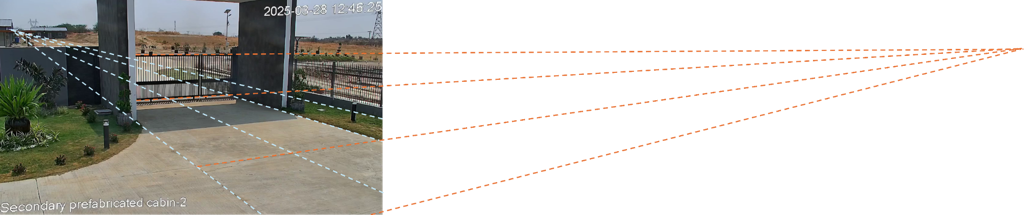}
    \caption{First Frame}
    \label{fig:calibrated_first}
  \end{subfigure}
  \vfill
  \begin{subfigure}[b]{0.9\textwidth}
    \centering
    \includegraphics[width=\textwidth]{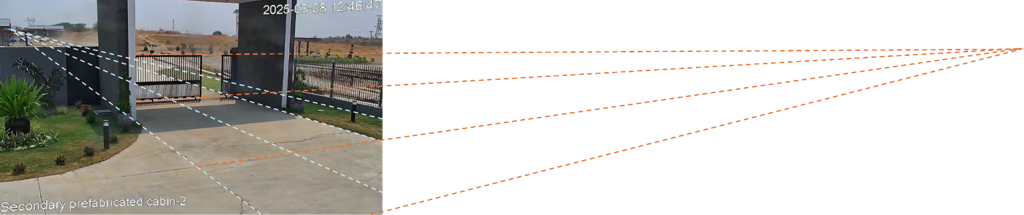}
    \caption{Last Frame}
    \label{fig:calibrated_last}
  \end{subfigure}
  \caption{Frames after applying stabilization and distortion correction.}
  \label{fig:calibration}
\end{figure}

\begin{figure}[htbp]  
\centering
\includegraphics[width=0.9\textwidth]{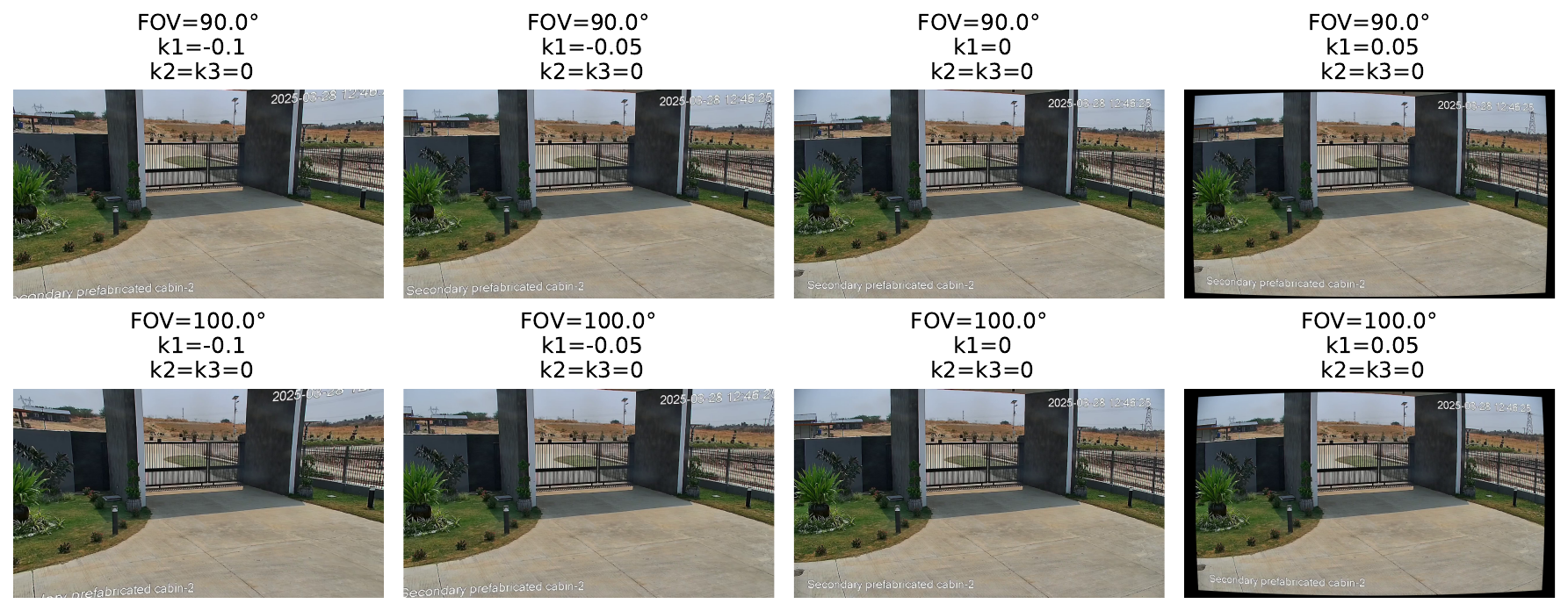}
\caption{Some other results with different sets of parameters.}
\label{fig:undistorted_images}
\end{figure}

In our case, we used $k_1 = -0.06$, $k_2 = 0.002$, and $k_3 = 0$, which produced visually aligned vanishing points between the first and last frames. 
This parameter set was selected based on its effectiveness in straightening architectural lines and aligning vanishing points.  The final calibrated results are shown in \autoref{fig:calibration}. After stabilization and calibration, the last frame shares the same vanishing points as the first frame. Results of some other parameter sets are also shown in \autoref{fig:undistorted_images}. Though there exists possibility that some other distortion parameter combinations of distortion parameters may yield close results, their impact on real-world displacement estimation remains limited—particularly when the measurement relies on the estimated dimensions of a reference object located near the region of interest. From this point of view, camera stabilization is necessary, while distortion correction is less demanded. Nevertheless, distortion correction is still applied in this research.


\subsection{Tracking Methods}\label{sec:tracking_method}

We applied computer vision–based multi-object tracking. Two tracking approaches were used: a traditional grayscale template matching method \cite{opencv_library} and a transformer-based model, CoTracker \cite{karaev23cotracker, karaev24cotracker3}. Then their results were validated through manual frame-by-frame annotation, though it cannot be taken as ground-truth because of the existence of human error.

Grayscale template matching was implemented using a traditional computer vision approach based on normalized cross-correlation applied to grayscale. Bounding boxes of interest were selected in the first frame, serving as templates to be tracked throughout the sequence. In each subsequent frame, the search for each template was restricted to a small window (4 pixels particularly) centered around its previous location to improve robustness against false matches. 

A state-of-the-art computer vision model, CoTracker3, was also employed for multi-point tracking. It has a transformer-based architecture designed for multi-object tracking. It jointly tracks multiple points or targets across frames and is capable of handling occlusions and reappearances. In this work, it was used to track selected points. The final provided time series in the appendix is based on it.

As illustrated in \autoref{fig:Boxes}, tracking-target objects were marked. All tracking results presented in \autoref{sec:result} correspond to the target indices shown in the figure. For all the three methods, pixel displacements were calculated as the relative distance of themselves to the initial frame.

\begin{figure}[htbp]  
\centering
\includegraphics[width=0.9\textwidth]{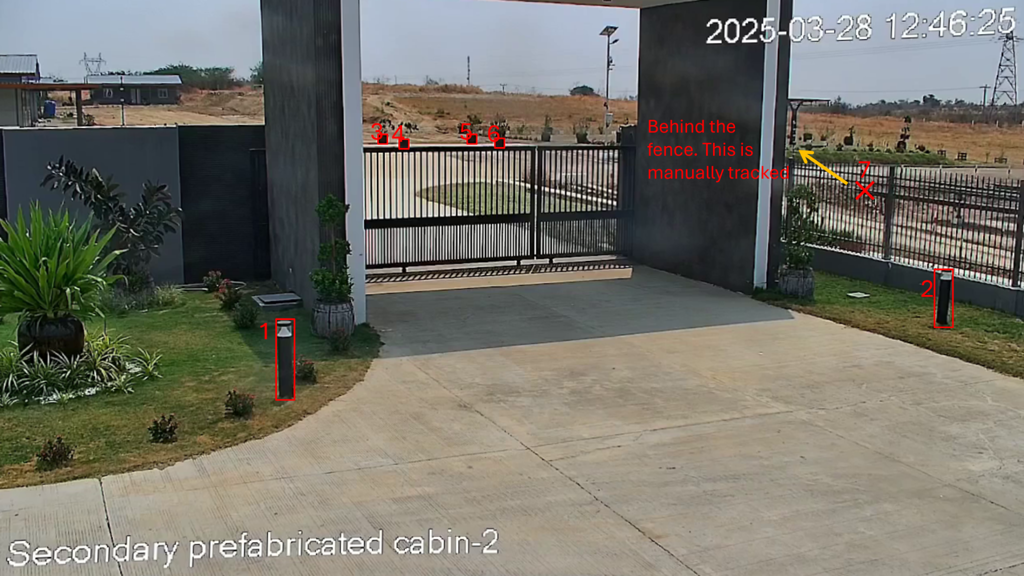}
\caption{Selected Targets. The results presented in \autoref{sec:result} correspond to the indices.}
\label{fig:Boxes}
\end{figure}

\section{Results} \label{sec:result}

\subsection{Pixel Displacements}\label{sec:result_pixel}
\begin{figure}[htbp]  
\centering
\includegraphics[width=0.9\textwidth]{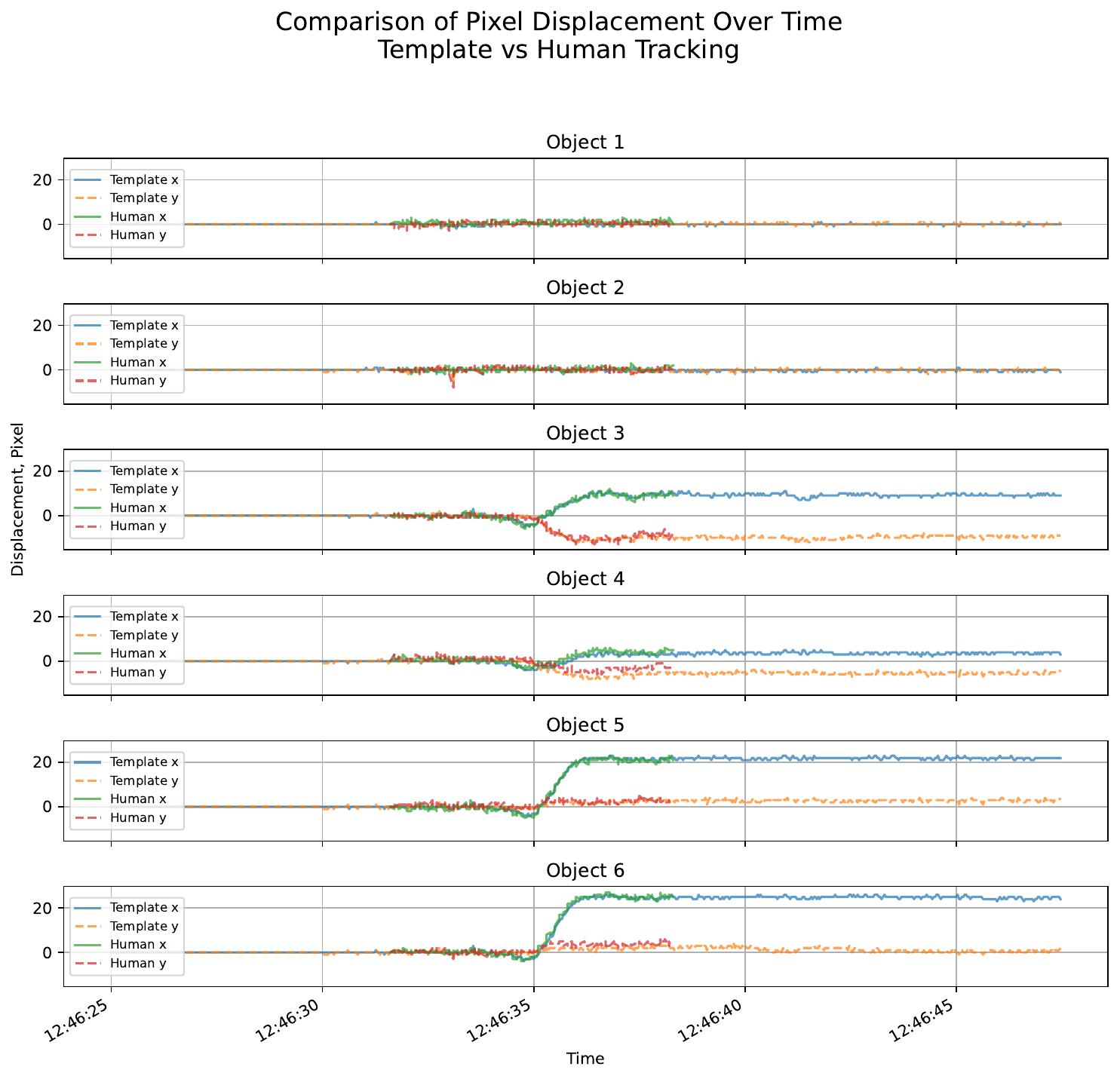}
\caption{Pixel displacement comparison: Template matching vs. Manual tracking.}
\label{fig:template_vs_human}
\end{figure}

\begin{figure}[htbp]  
\centering
\includegraphics[width=0.9\textwidth]{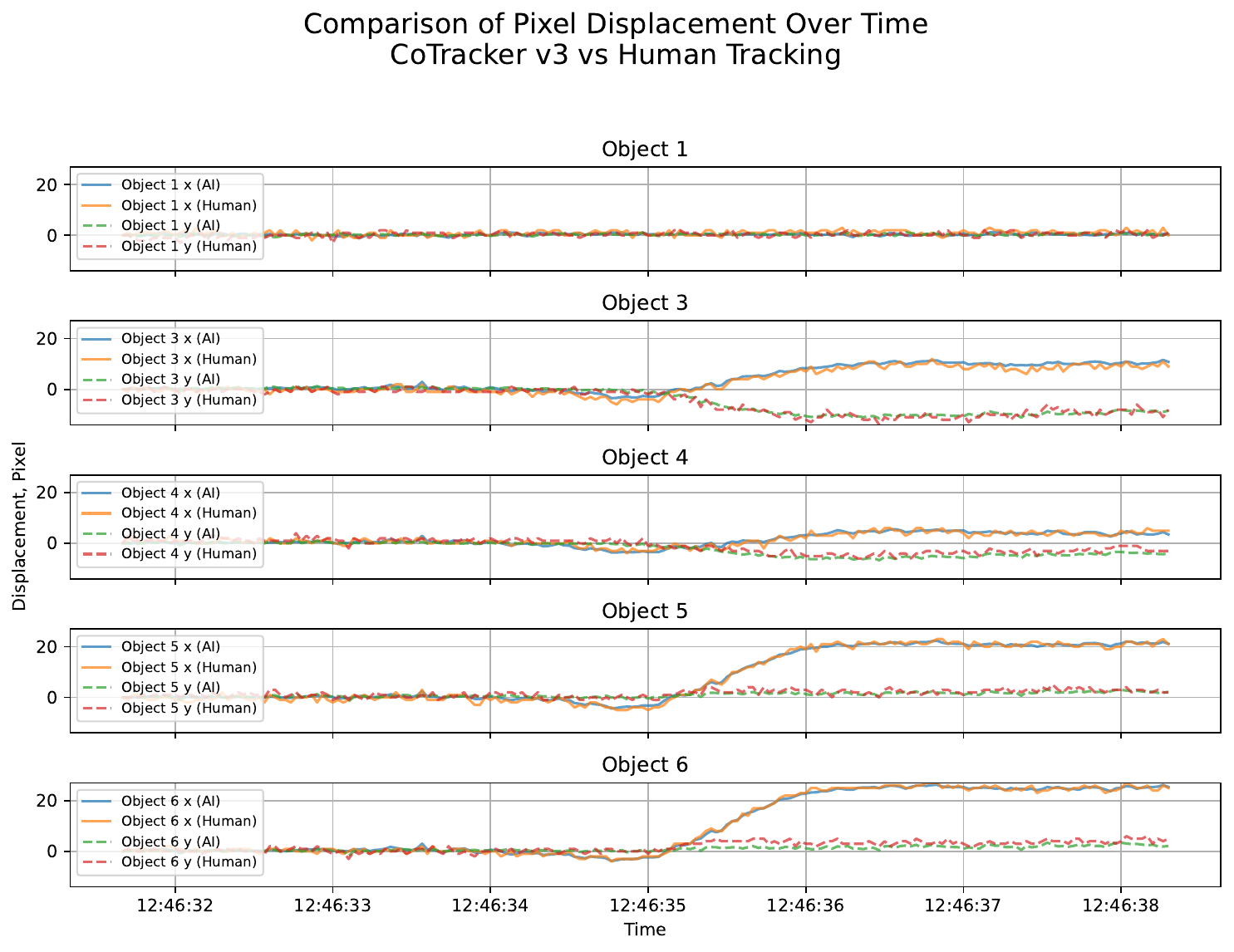}
\caption{Pixel displacement comparison: CoTracker vs. Manual tracking.}
\label{fig:ai_vs_human}
\end{figure}

\begin{figure}[htbp]
\centering
\includegraphics[width=0.9\textwidth]{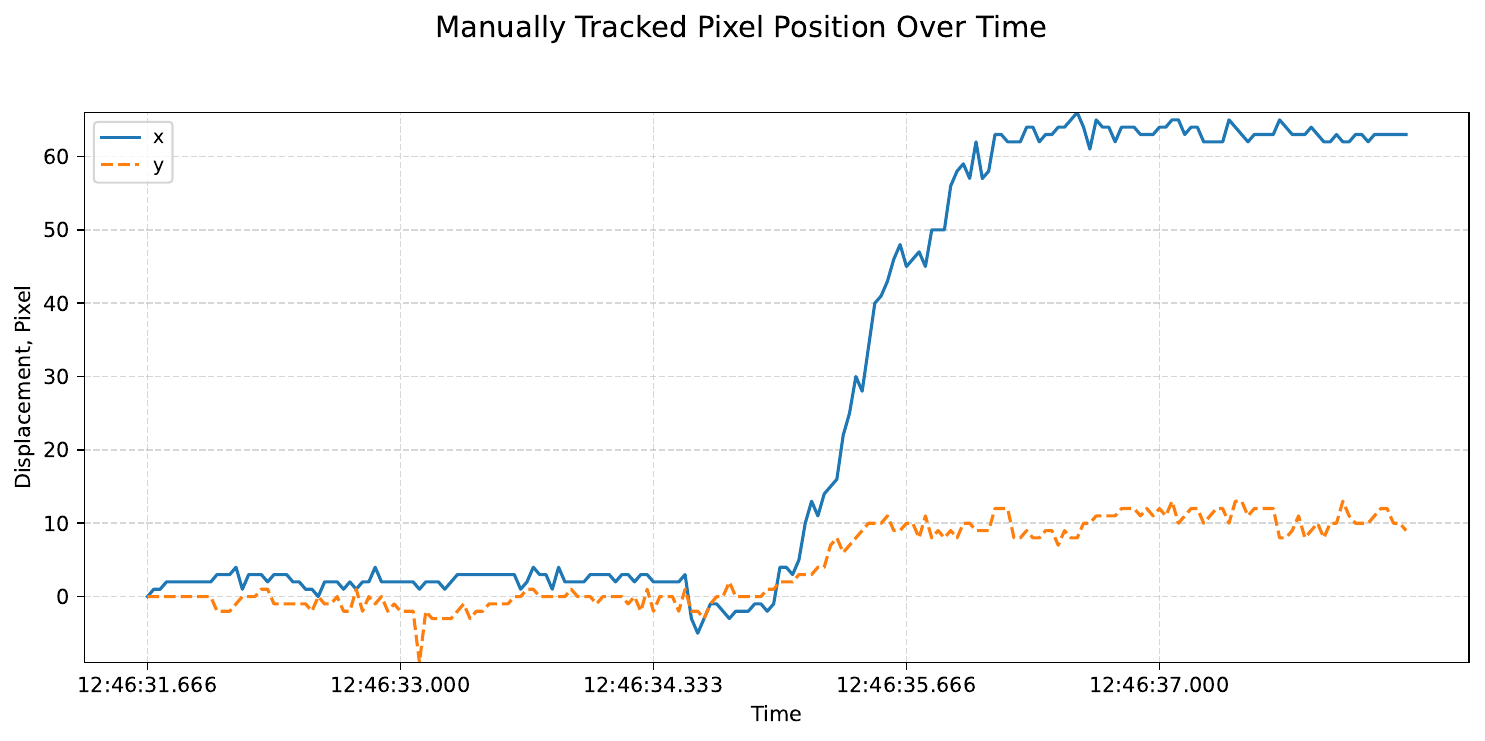}
\caption{Pixel displacement of object 7.}
\label{fig:Fence_Track}
\end{figure}

Comparisons were conducted between manual and automated tracking techniques. Template matching and Cotracker-based methods were compared against manual tracking, as shown in \autoref{fig:template_vs_human} and \autoref{fig:ai_vs_human}. It can be seen that, except for object 4 in \autoref{fig:template_vs_human}, which is partially occluded by the top edge of the gate, all results are well aligned with each other. This provides confidence in the tracking results. Additionally, object 7, which is located behind the fence, is tracked manually because none of the tested computer vision methods performed satisfactorily. The result is shown in \autoref{fig:Fence_Track}. It should be noted that the CCTV footage time is different than the actual time of MST. All results in this article use the CCTV time from its video watermark.

\subsection{Physical Displacements}\label{sec:physical}

\subsubsection{Estimation from Plant Pot}

\begin{figure}[htbp]
\centering
\begin{subfigure}[b]{0.48\textwidth}
    \centering
    \includegraphics[width=\textwidth]{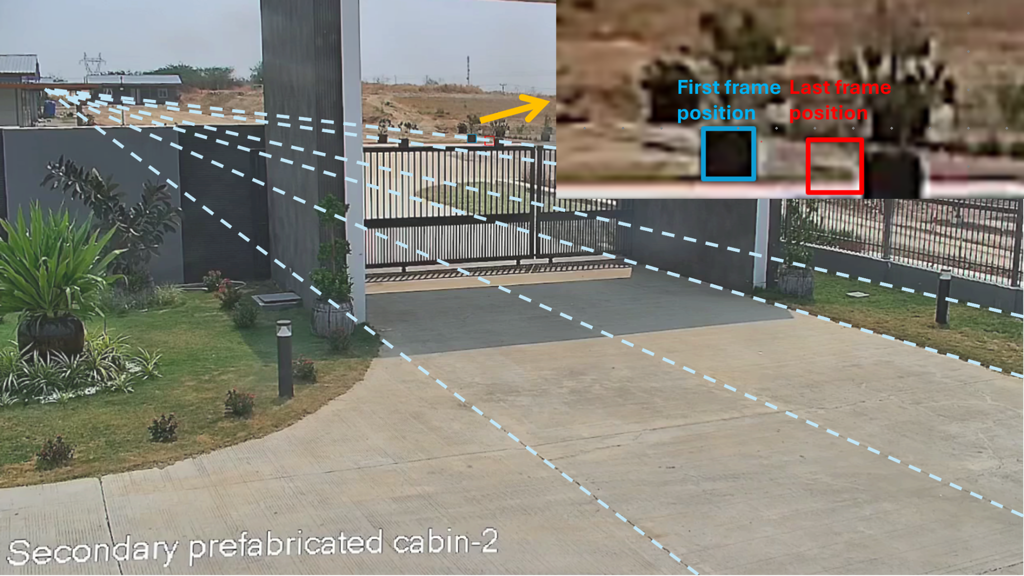}
    \caption{Displacement of the target object.}
\end{subfigure}
\hfill
\begin{subfigure}[b]{0.48\textwidth}
    \centering
    \includegraphics[width=\textwidth]{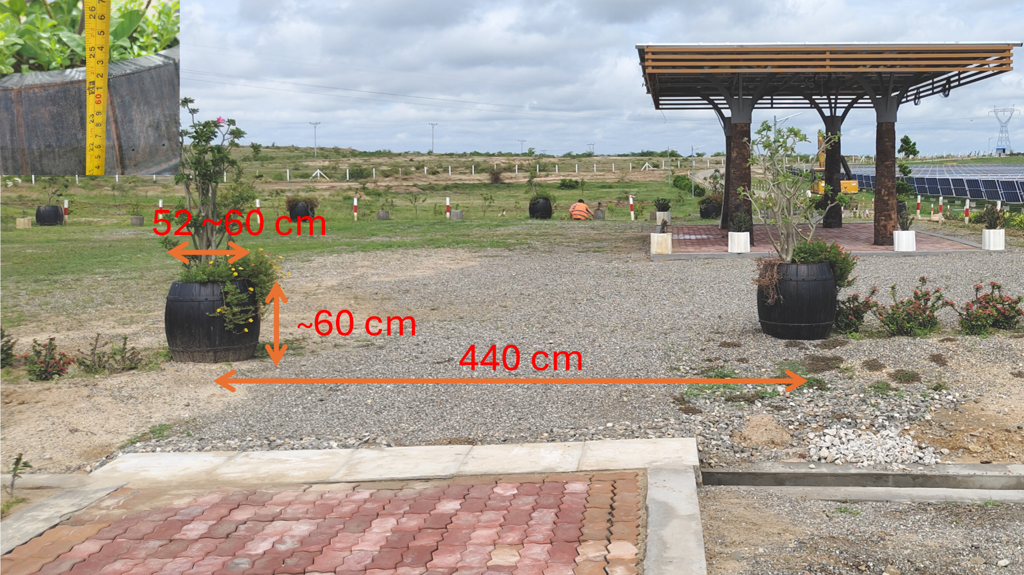}
    \caption{Physical dimensions of the plant pot.}
\end{subfigure}
\caption{Displacement and physical dimensions of the target object.}
\label{fig:Physical_Target_Movement}
\end{figure}

As shown in \autoref{fig:Physical_Target_Movement}, the corresponding physical dimensions of plant pots were obtained on-site. Accordingly, the offset is estimated to be approximately 280 cm based on linear interpolation of the distance between the two plant pots. Note that the current distance between the two plant pots is not guaranteed to be identical to that pair at the time of the earthquake, but is assumed to be close based on discussions with field staffs. 

\subsubsection{Estimation from the Fence}

\begin{figure}[htbp]
\centering
\includegraphics[width=0.9\textwidth]{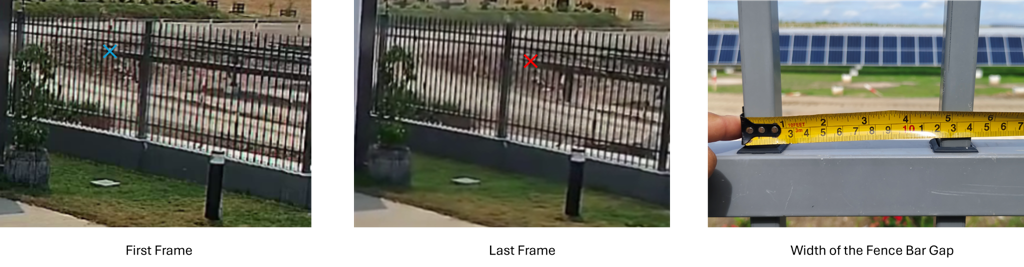}
\caption{Tracking solar panel behind fence.}
\label{fig:Distance_FirstLast_Fence}
\end{figure}

As shown in \autoref{fig:Distance_FirstLast_Fence}, within the camera’s field of view, the fence on the right side consists of evenly spaced vertical bars, providing a fixed physical reference grid. Several distinct target objects behind the fence—such as the columns of a pavilion and the support post of a solar panel on the right side of the road—are visible through the gaps. As these objects moved during the rupture, their relative positions can be tracked over time based on the shifting projections on the fence grid. By analyzing the motion of these projections along the fence and applying the principle of similar triangles, the actual displacements of the target objects can be estimated. Specifically, using the known distances between the camera and the fence, and between the fence and the target objects, we obtain the displacement of the upper-left corner of the solar panel.

\begin{figure}[htbp]
\centering
\includegraphics[width=0.9\textwidth]{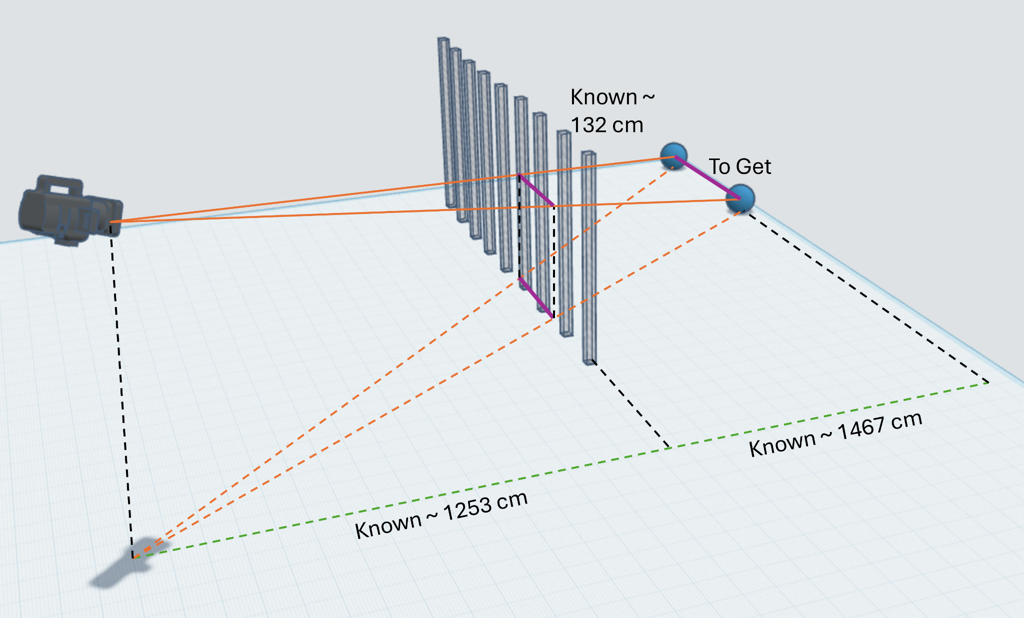}
\caption{Estimation from fence bar gap. This diagram was created using Tinkercad, a free online design tool by Autodesk. The CAD geometry is only for conceptual illustration, not exact.}
\label{fig:Fence_Illustration}
\end{figure}

Using the simple projection relationship shown in \autoref{fig:Fence_Illustration}, we inferred a displacement parallel to the fence of approximately 287 cm for the upper-left corner of the solar panel. Using this value as a reference, the pixel displacement time series in \autoref{fig:Fence_Track} can be converted  into a physical displacement time series.


\subsubsection{Direct Measurement of Curb}

\begin{figure}[htbp]
\centering
\includegraphics[width=0.9\textwidth]{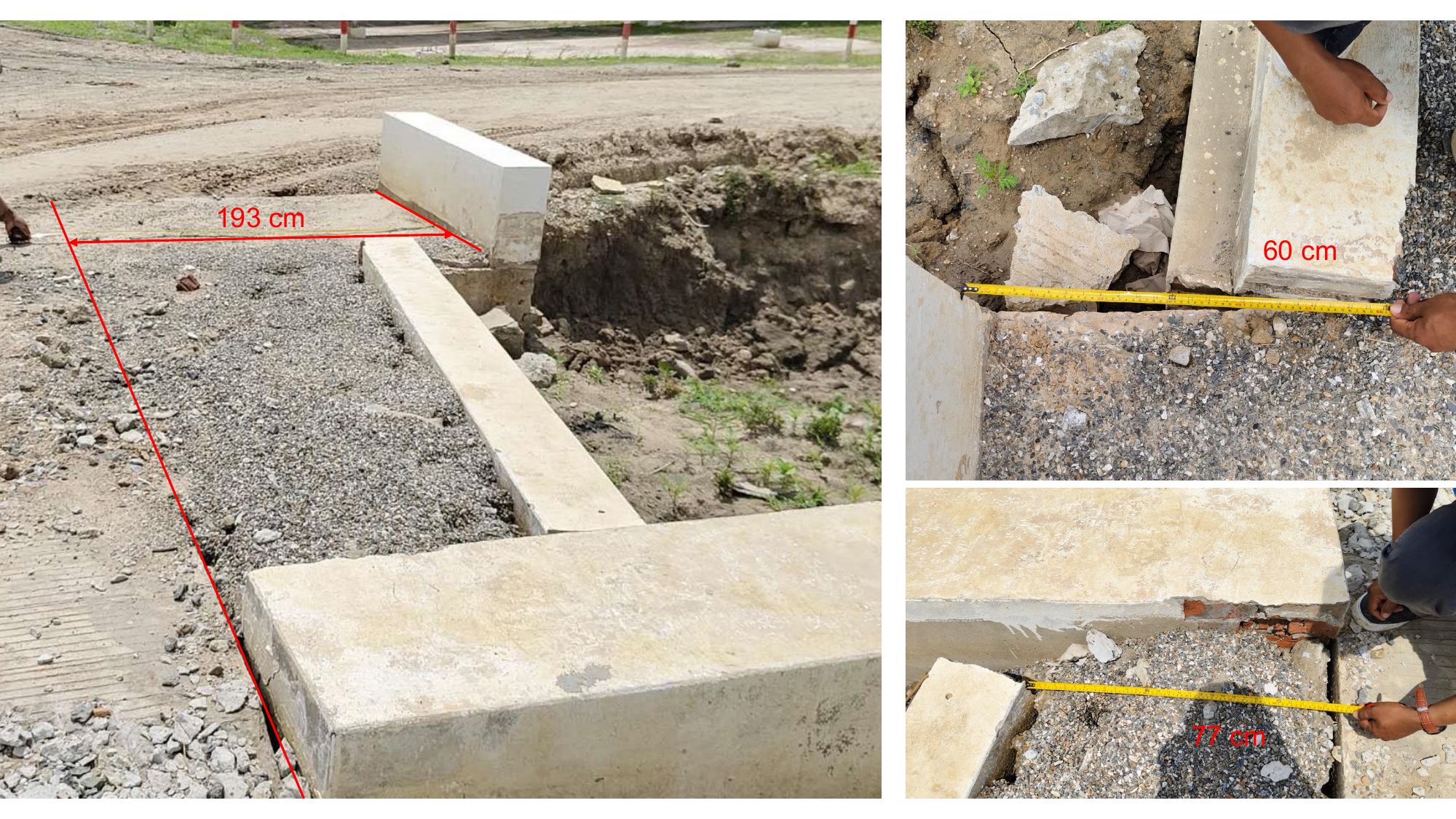}
\caption{Direct Measurement of Fault Relative Displacement.}
\label{fig:Direct_Fault}
\end{figure}

In the area recorded by the surveillance camera, there is a section of concrete curb located just to the right of the main gate that crosses the fault trace. As shown in \autoref{fig:Direct_Fault}, we conducted a field measurement of the offset, which represents the cumulative coseismic slip at this specific location. The measured relative displacement of concrete curb is 193 cm. This value is smaller than the previously obtained measurements of approximately 280 and 287 cm. The discrepancy arises primarily because the current measurement uses the near-side concrete curb as a reference point, which itself experienced coseismic displacement. Additionally, the structural rigidity of the structure have partially resisted the movement. Consequently, the measured value reflects the relative slip of the curb structure and is, as expected, smaller than the actual slip. Therefore, the lower limit of the ground slip is believed to be at least 193 cm.

\subsubsection{Estimated Physical Displacement Time Series}

From the pixel displacement results presented in \autoref{sec:result_pixel}, we can see objects 1 and 2 show minimal displacement as intended, due to the video stabilization applied.
From the video, it is evident that object 3 exhibits relatively large anti-ground movement, deviating from the original ground plane, and is therefore excluded from physical displacement estimation. The black coloration of objects 4 and 6 makes them difficult to distinguish from the gate, which may reduce the robustness of the algorithm results. The manually tracking result of solar panel (object 7) is poor. Consequently, the results obtained from object 5 is the more reliable among all objects, and the result of it is presented here.

\begin{figure}[htbp]
\centering
\includegraphics[width=0.9\textwidth]{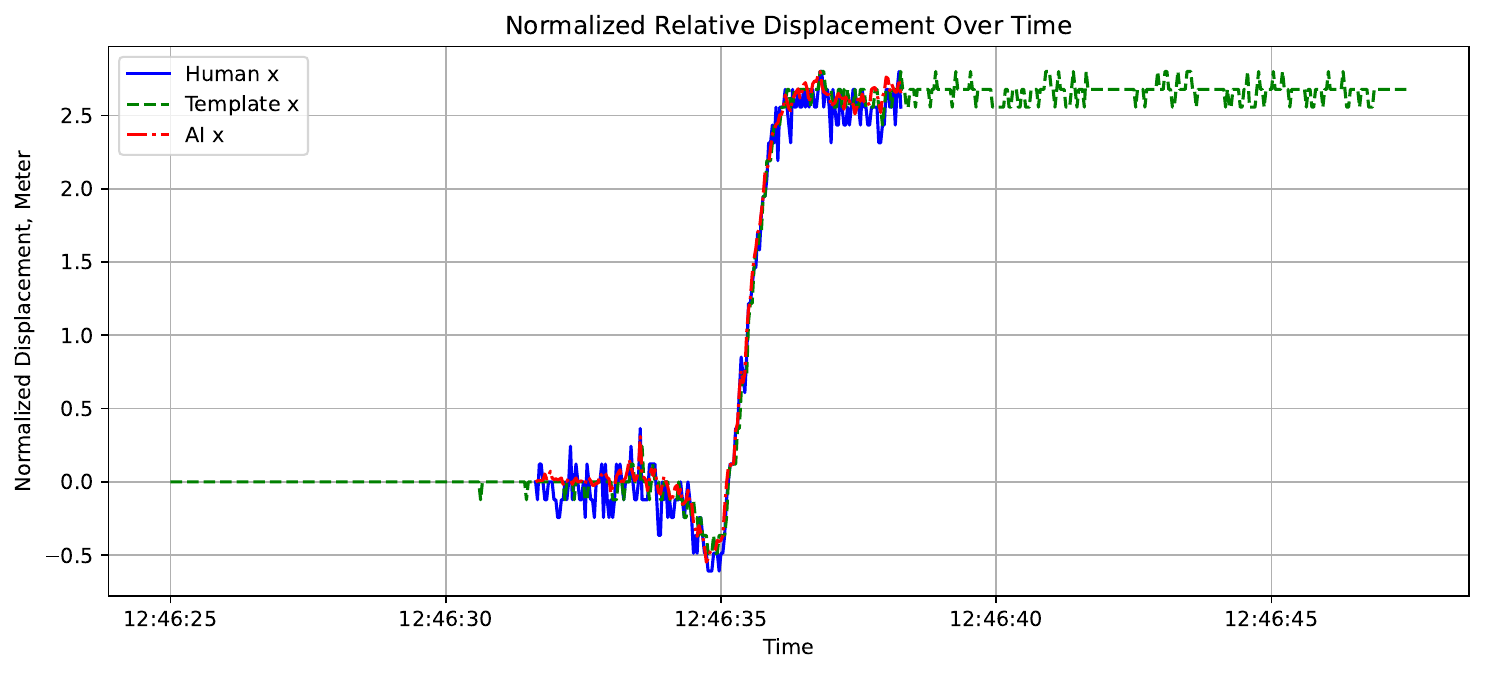}
\caption{Real-work displacement estimation of the plant pot.}
\label{fig:physical displacement}
\end{figure}

The estimated offset 280 cm is then used to normalize each x-displacement series to obtain physical displacement time series, as shown in \autoref{fig:physical displacement}.


\subsection{Velocity Range and Critical Slip-Weakening Distance (\(D_c\))}

\begin{figure}[htbp]
\centering
\includegraphics[width=0.9\textwidth]{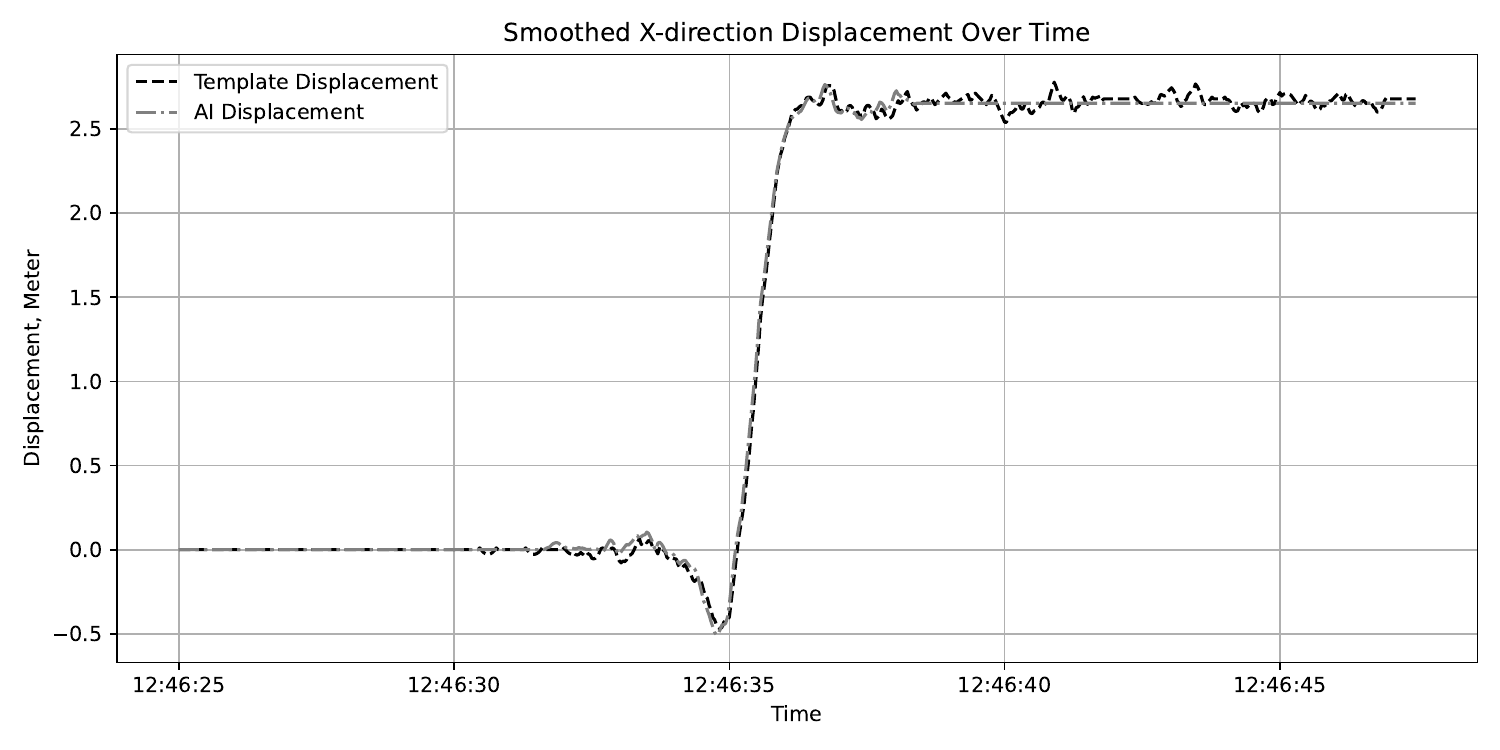}
\caption{Smoothed displacement time series derived from template matching and CoTracker tracking.}
\label{fig:Smoothed Displacement}
\end{figure}

Based on the displacement fields obtained from template matching and CoTracker-based tracking, we analyzed the temporal evolution of fault-parallel ground motion. To mitigate differentiation instability caused by limited spatial resolution while preserving the overall motion characteristics, we applied a Savitzky-Golay filter \cite{chen2004simple} to each displacement-time series. The smoothed results are shown in Figure~\autoref{fig:Smoothed Displacement}.

Velocity time series were derived from the smoothed displacement signals via central differentiation. The resulting velocity profiles show an initial negative phase followed by rapid positive acceleration and deceleration, indicating the existence of a reversal slip before the onset of dynamic rupture, which may be effected by the seismic waves propagated through the medium prior to fault instability, a phenomenon also observed in near-fault strong motion records during the 2023 Turkey earthquake \cite{ding2023sharp}.

\begin{figure}[htbp]
\centering
\includegraphics[width=0.9\textwidth]{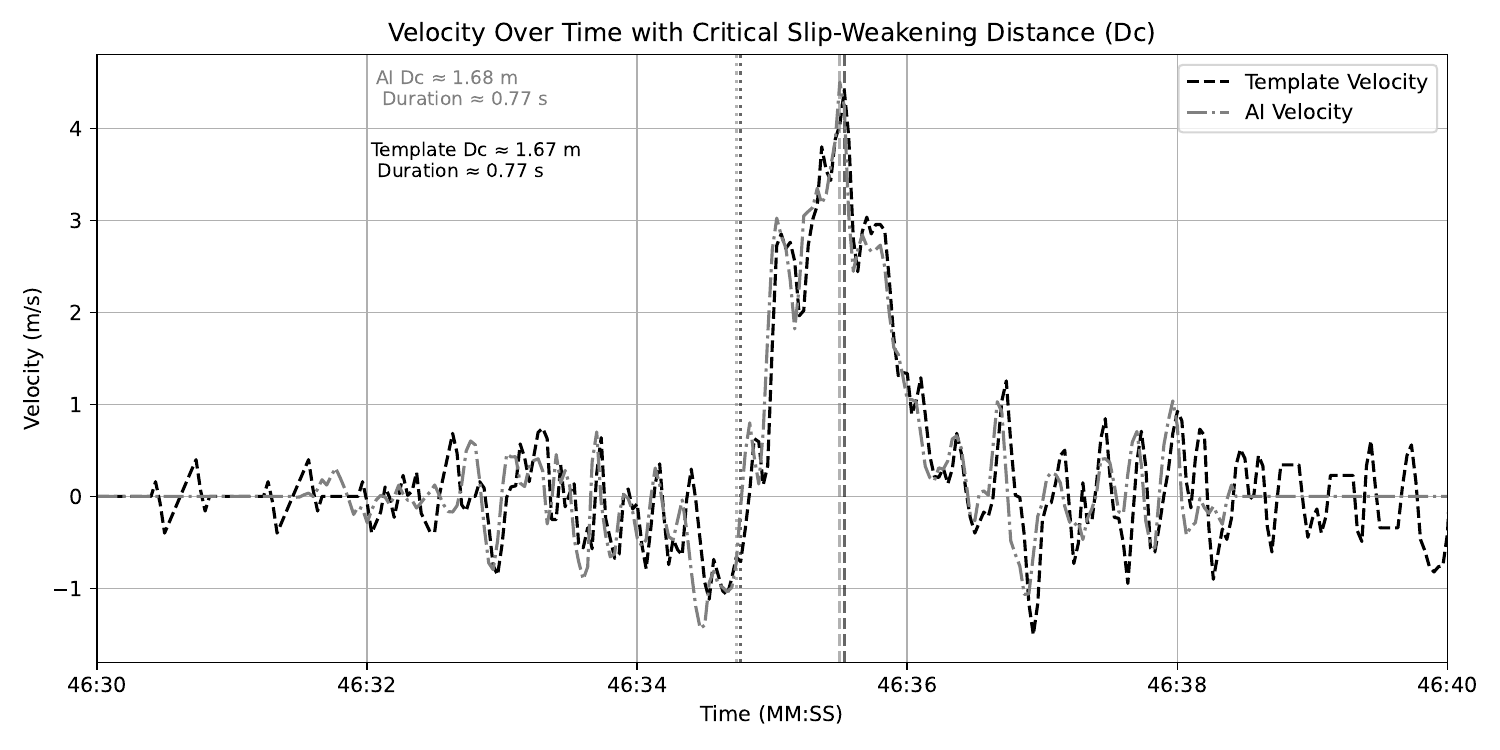}
\caption{Velocity profiles and estimated critical slip-weakening distances ($D_c$) derived from different tracking methods.}
\label{fig:Dc_calculation}
\end{figure}

To estimate the critical slip-weakening distance (\(D_c\)), we identify the point of minimum displacement as the onset of dynamic rupture and the subsequent peak in velocity as the end of the window. Dc is calculated as the integral of velocity over this interval using Simpson’s rule. The resulting velocity curves and Dc values are visualized and annotated in \autoref{fig:Dc_calculation}, providing comparisons of dynamic slip behavior across different motion-tracking methods.

\section{Discussion}\label{sec:discussion}
In this research, we applied a machine vision method, along with field measurement, to obtain the displacement time series of the coseismic surface rupture during the 2025 Mw7.7 Myanmar earthquake. 
Our results demonstrate that computer vision methods provide displacement time series consistent with manual tracking. Moreover, the on-fault displacements inferred from different sources are in good agreement, supporting a certain level of reliability in displacement measurements. The slips estimated from the plant pot and the solar panel corner are both approximately 280 cm, whereas the 193 cm displacement of the curb represents a lower-bound estimate.

\begin{figure}[htbp]
\centering
\includegraphics[width=0.9\textwidth]{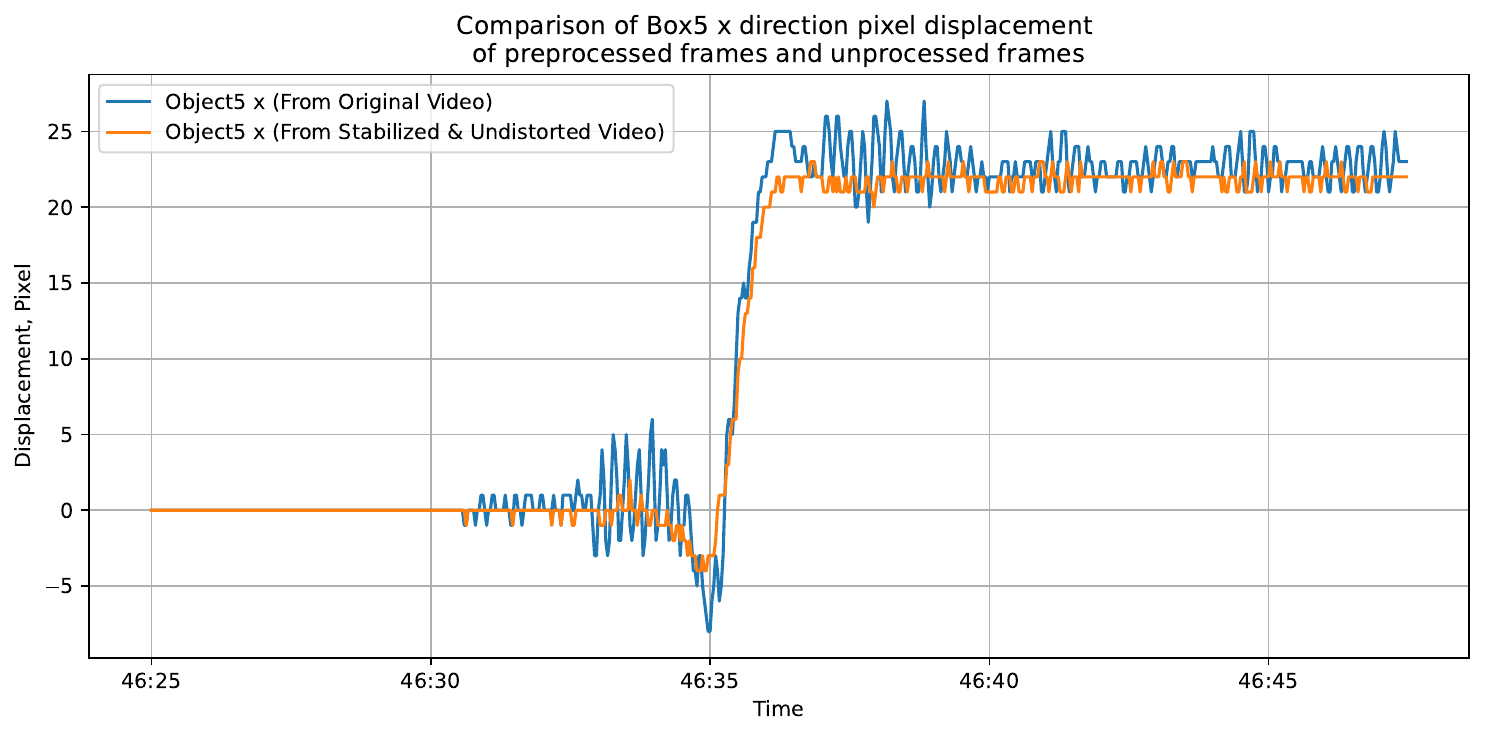}
\caption{Comparison of tracking results between the original video and the calibrated video.}
\label{fig:comparison_box5_x}
\end{figure}

However, the final displacement time series involves several sources of uncertainty. The primary source is the projection estimation based on field-measured references. Such a result should be regarded as an approximation. A second major source of error arises from camera shaking and lens distortion. As shown in \autoref{fig:comparison_box5_x}, though the extent of the remaining uncalibrated error remains unclear, much of the small vibration is effectively canceled after stabilization and distortion correction. The effect of distortion correction is limited in this study.

\begin{figure}[htbp]
\centering
\includegraphics[width=0.9\textwidth]{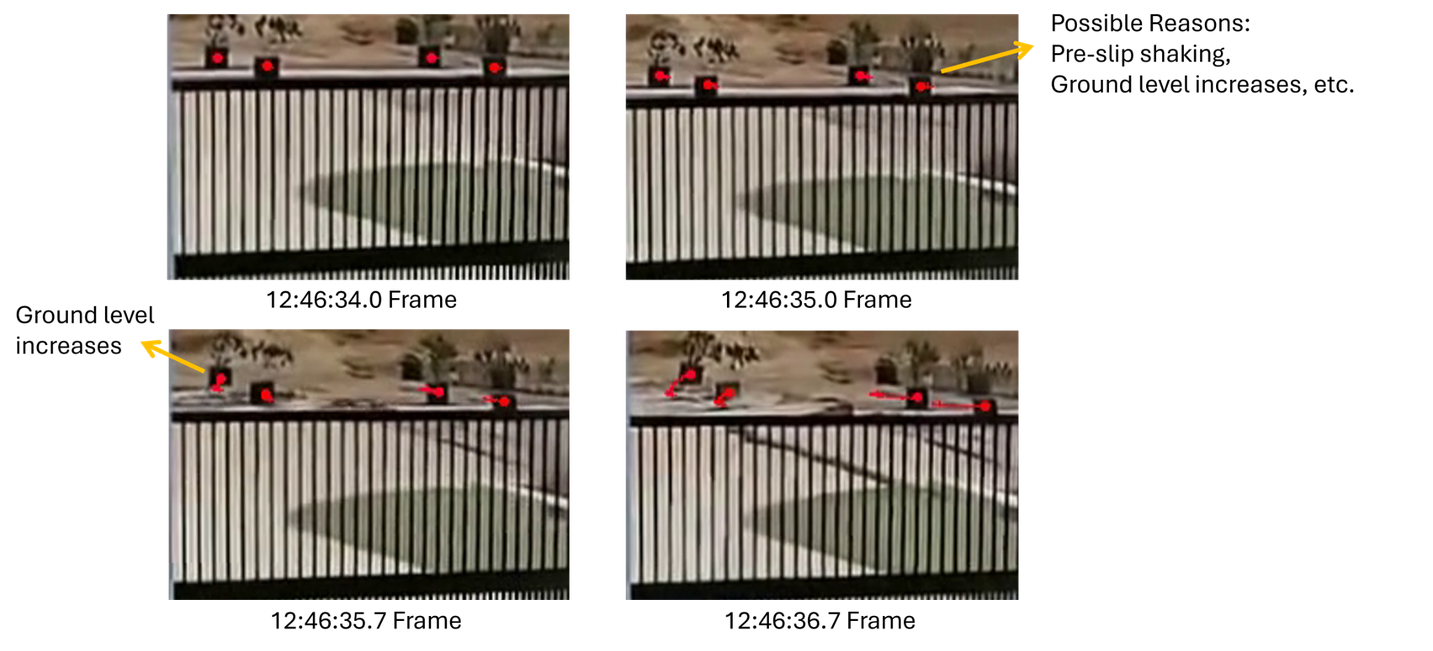}
\caption{Deformation occurs in the anti-slip direction prior to the slip, and some regions suffer an increase in ground elevation. Object tracking paths are obtained by CoTracker model.}
\label{fig:Ground_Rise}
\end{figure}

It can be noted that there exists an anti-slip deformation before the start of the slip. 
It can be directly observed from timestamp 12:46:34 to 12:46:35 as shown in \autoref{fig:Ground_Rise}.
The reason could be the fault underwent unstable slip, seismic waves were already propagating. This phenomenon is also observed in \cite{latour2025directestimationearthquakesource} for a different tracked object using different method (See their Figure 3(c)). This observation has also been reported in Turkey 2023 Mw 7.8 earthquakes \cite{ding2023sharp}.

As the offset given from satellite \cite{van_wyk_de_vries_2025_15123647}, the slip at this location is approximately 4-5 meters, which differs from the value inferred from the video. Possible reason is the local surface rupture displacement does not fully represent the slip in the fault zone. The estimation from \cite{latour2025directestimationearthquakesource} is also smaller than the satellite observation.

Besides, we noted that the rupture and ground motion is nearly parallel to the boundary between concrete-paved ground and adjacent soil ground, as shown in \autoref{fig:Measure2}, suggesting the shear slip path may possibly be influenced by the pavement foundation reinforcement. In crack path analysis, this factor should be treated with caution.

Finally, it is important to re-emphasize that, unless otherwise specified, the timestamps provided throughout the results are based on the video watermark and do not reflect the actual time at which the wavefront reaches the camera location. In fact, the video footage time is approximately 4 minutes behind MST.




\section{Conclusion}

A recent video footage captures the on-fault rupture process during the Myanmar earthquake, offering a unique observational time series data of the along-strike slip. We present the technical details and results of extracting the coseismic surface displacement time series from this footage. The workflow involves video stabilization, distortion correction, object tracking, and scale conversion. Two computer vision-based object tracking methods were employed: CoTracker and a template matching algorithm. The results were manually validated to ensure that object motion was reliably tracked in pixel space.


This on-fault rupture observation marks the very first time that a continuous slip time series has been directly measured in-situ during an earthquake. The data enables researchers to quantitatively validate dynamic rupture models against real‐world data, refine the parameters governing fault friction and slip evolution, and benchmark seismic hazard assessments with unprecedented precision. In effect, this dataset opens the door to re-examining long-standing assumptions about rupture propagation, calibrating ground-motion simulators, and enhance our understanding and modeling of future seismic scenarios.

\section*{Acknowledgement}

We sincerely thank Green Power Energy Co., Ltd. (Myanmar) and Sungrow Power Supply Co., Ltd. (Hefei, China) for generously providing both the video footage capturing the coseismic rupture process and the on-site measurements, which were invaluable to this study. In particular, we are deeply grateful to Yao Meng, James, Tony Yang, Kyaw Zin Htet, and Ko Aung Kyaw Pyae Phyo for their support in supplying first-hand field data. Their contributions are gratefully acknowledged. All software packages used in this study were licensed under open-source agreements.

\clearpage
\section*{Appendix I: Field Measurement}\label{sec:appendix_I}


\begin{figure}[htbp]  
\centering
\includegraphics[width=1.0\textwidth]{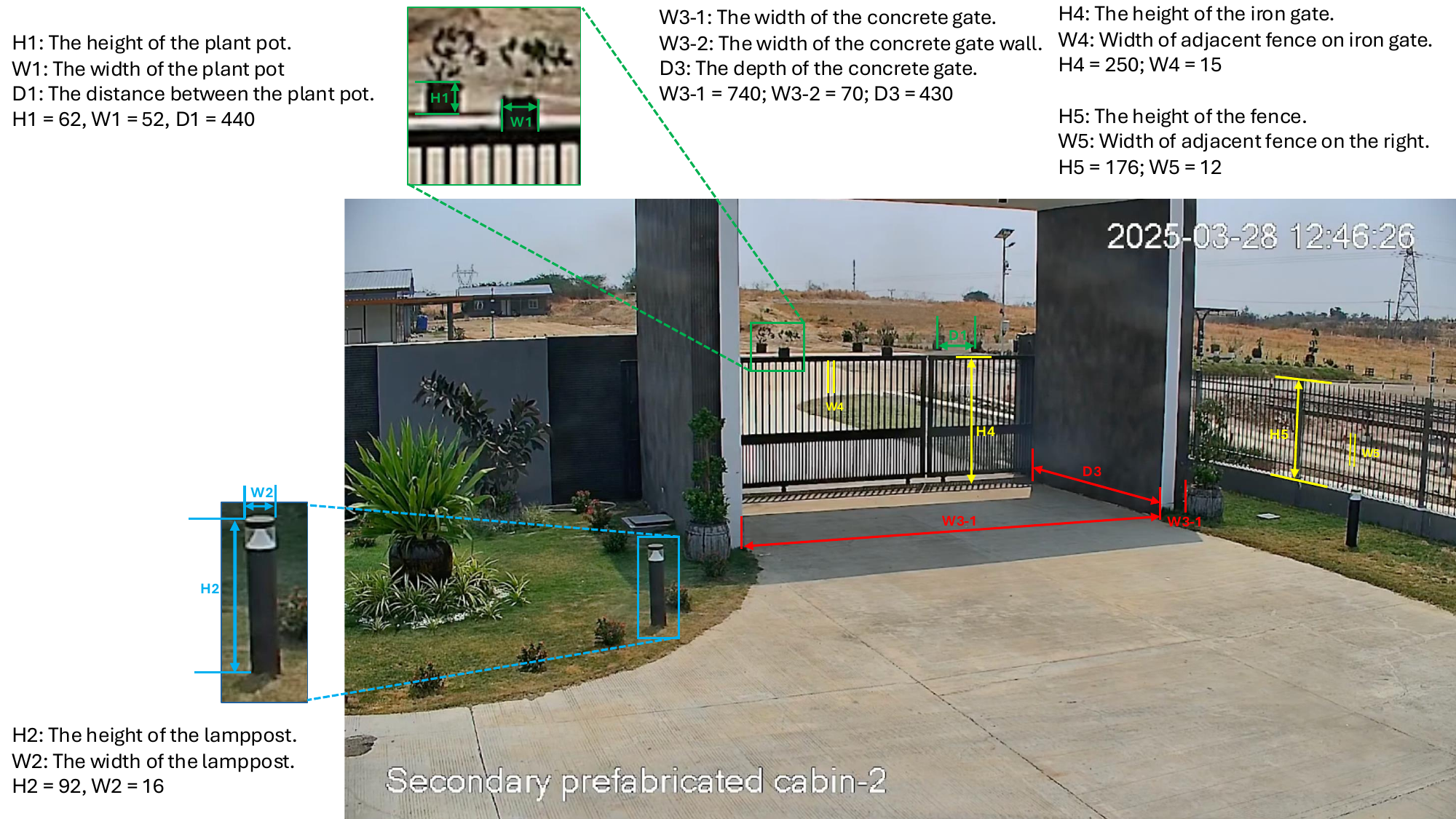}
\caption{Reference Object Size for Video Analysis. Measured in cm.}
\label{fig:Measure1}
\end{figure}

\begin{figure}[htbp]  
\centering
\includegraphics[width=0.9\textwidth]{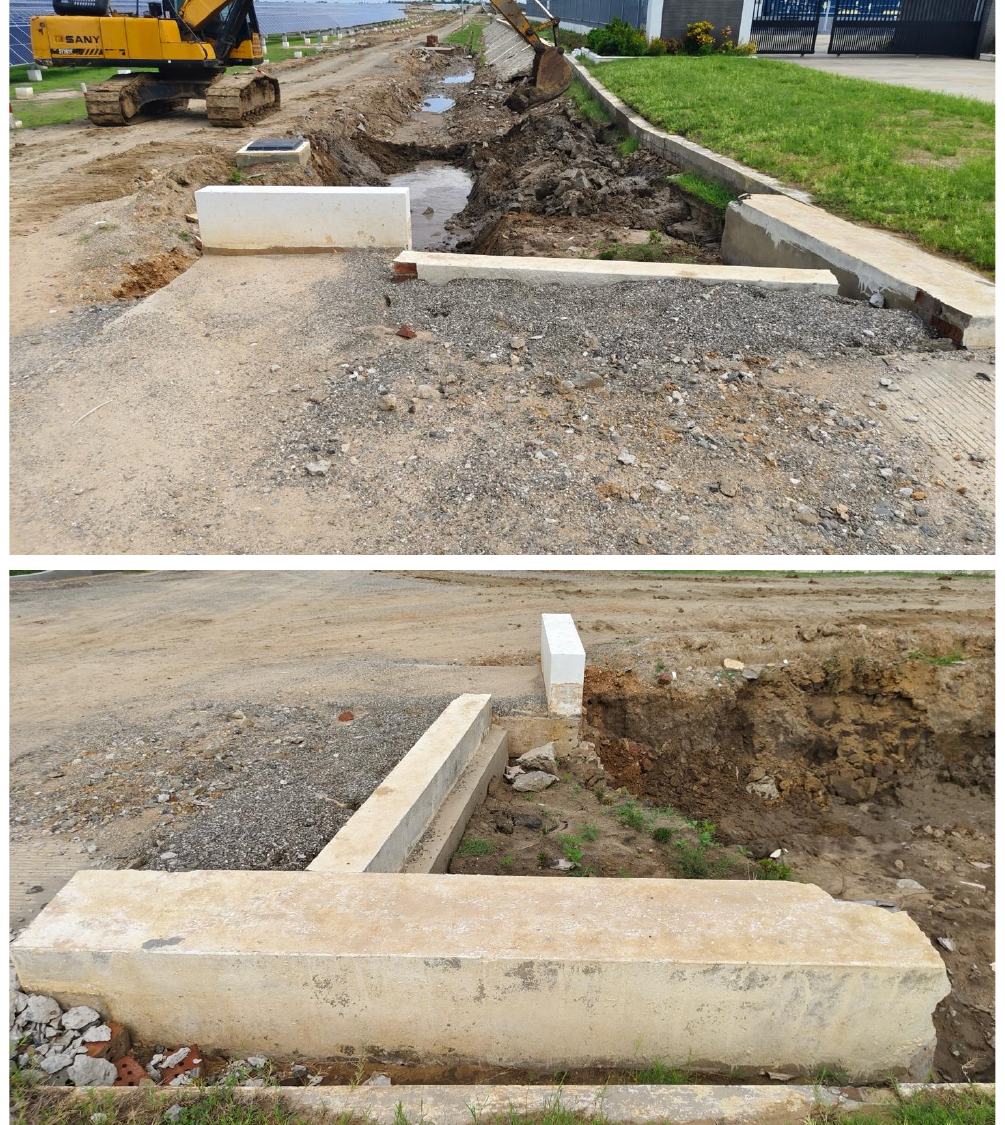}
\caption{Direct observation of fault displacement after earthquake.}
\label{fig:Measure2}
\end{figure}

\clearpage
\section*{Appendix II: Displacement Time-Series}\label{sec:appendix_II}

For the length limitation, here in \autoref{tab:PixelDis_Cotracker}, we only include the CoTracker result of the object 5. The sub-pixel precision is inferred by CoTracker model. Note that the result presented here may have negligible difference from those shown in the preceding article figures due to the ad hoc selection of tracking regions and numerical round up.

\begin{table}[htbp]
    \centering
    \begin{tabular}{|l l l|l|l l l|l|l l l|l|l l l|l|l l l|}
    \hline
        time & x & y & ~ & time & x & y & ~ & time & x & y & ~ & time & x & y  \\ \hline
        0.000 & 0.00 & 0.00 & ~ & 0.633 & 0.14 & -0.04 & ~ & 1.267 & 0.45 & 0.57 & ~ & 1.900 & 2.47 & 0.79  \\ \hline
        0.033 & 0.06 & 0.08 & ~ & 0.667 & -0.13 & -0.10 & ~ & 1.300 & 0.24 & 0.68 & ~ & 1.933 & 0.26 & 0.14  \\ \hline
        0.067 & 0.06 & 0.05 & ~ & 0.700 & 0.08 & 0.34 & ~ & 1.333 & 0.15 & 0.25 & ~ & 1.967 & 0.25 & 0.98  \\ \hline
        0.100 & 0.08 & 0.02 & ~ & 0.733 & 0.21 & 0.41 & ~ & 1.367 & -0.62 & 0.47 & ~ & 2.000 & 0.11 & 0.44  \\ \hline
        0.133 & 0.09 & 0.06 & ~ & 0.767 & 0.00 & 0.42 & ~ & 1.400 & -0.26 & 0.33 & ~ & 2.033 & -0.48 & 0.09  \\ \hline
        0.167 & 0.43 & 0.18 & ~ & 0.800 & 0.02 & 0.21 & ~ & 1.433 & -0.28 & -1.57 & ~ & 2.067 & 0.16 & 0.24  \\ \hline
        0.200 & 0.16 & 0.18 & ~ & 0.833 & 0.06 & 0.14 & ~ & 1.467 & 0.28 & 0.15 & ~ & 2.100 & 0.71 & 0.34  \\ \hline
        0.233 & 0.29 & 0.23 & ~ & 0.867 & -0.10 & 0.19 & ~ & 1.500 & 0.50 & 0.06 & ~ & 2.133 & 0.34 & 0.32  \\ \hline
        0.267 & 0.58 & 0.68 & ~ & 0.900 & -0.12 & 0.11 & ~ & 1.533 & 0.66 & 0.25 & ~ & 2.167 & 0.67 & 0.44  \\ \hline
        0.300 & 0.21 & 0.22 & ~ & 0.933 & 0.25 & 0.88 & ~ & 1.567 & -0.18 & 0.03 & ~ & 2.200 & 0.00 & 0.51  \\ \hline
        0.333 & 0.13 & -0.10 & ~ & 0.967 & 0.21 & 0.84 & ~ & 1.600 & 0.09 & 0.10 & ~ & 2.233 & -0.57 & 0.61  \\ \hline
        0.367 & 0.17 & -0.93 & ~ & 1.000 & 0.05 & -0.02 & ~ & 1.633 & 0.21 & 0.07 & ~ & 2.267 & -0.67 & 0.17  \\ \hline
        0.400 & 0.11 & -0.02 & ~ & 1.033 & -0.19 & 0.11 & ~ & 1.667 & 0.63 & 0.49 & ~ & 2.300 & -0.13 & 0.46  \\ \hline
        0.433 & 0.15 & 0.05 & ~ & 1.067 & -0.18 & 0.16 & ~ & 1.700 & 1.21 & 0.55 & ~ & 2.333 & 0.31 & 0.55  \\ \hline
        0.467 & 0.21 & 0.20 & ~ & 1.100 & -0.06 & 1.65 & ~ & 1.733 & 0.79 & 0.22 & ~ & 2.367 & 0.02 & 0.31  \\ \hline
        0.500 & -0.13 & 0.16 & ~ & 1.133 & -0.08 & 0.07 & ~ & 1.767 & 0.45 & 0.89 & ~ & 2.400 & -0.26 & 0.45  \\ \hline
        0.533 & 0.06 & 0.02 & ~ & 1.167 & 0.43 & 0.49 & ~ & 1.800 & 0.07 & 0.93 & ~ & 2.433 & -0.77 & 0.81  \\ \hline
        0.567 & 0.02 & 0.64 & ~ & 1.200 & 0.36 & 0.56 & ~ & 1.833 & 0.18 & 0.62 & ~ & 2.467 & -1.06 & 0.87  \\ \hline
        0.600 & 0.21 & 0.56 & ~ & 1.233 & 0.54 & 1.23 & ~ & 1.867 & 0.70 & 0.41 & ~ & 2.500 & -0.73 & 0.29  \\ \hline
    \end{tabular}
    \caption{Pixel displacement time series obtained by CoTracker. Time unit is second, starting from 12:46:31.666 PM video watermark time. Displacement unit is pixel. To obtain a real-world displacement estimation, a suggested normalization factor of maximum displacement 280 cm is recommended based on on-site survey.}
    \label{tab:PixelDis_Cotracker}
\end{table}

\begin{table}[!ht]
    \centering
    \begin{tabular}{|l l l|l|l l l|l|l l l|l|l l l|l|l l l|}
    \hline
        time & x & y & ~ & time & x & y & ~ & time & x & y & ~ & time & x & y  \\ \hline
        2.533 & -0.29 & 0.45 & ~ & 3.167 & -3.83 & 0.23 & ~ & 3.800 & 5.90 & 1.42 & ~ & 4.433 & 19.96 & 1.30  \\ \hline
        2.567 & -0.18 & 0.38 & ~ & 3.200 & -3.68 & -0.74 & ~ & 3.833 & 8.08 & 1.55 & ~ & 4.467 & 19.99 & 2.03  \\ \hline
        2.600 & -0.65 & -0.12 & ~ & 3.233 & -3.74 & -0.16 & ~ & 3.867 & 9.39 & 1.60 & ~ & 4.500 & 20.35 & 1.18  \\ \hline
        2.633 & -0.40 & 0.57 & ~ & 3.267 & -3.66 & -0.09 & ~ & 3.900 & 9.99 & 1.63 & ~ & 4.533 & 21.05 & 2.00  \\ \hline
        2.667 & -1.22 & 0.43 & ~ & 3.300 & -3.18 & 0.34 & ~ & 3.933 & 11.28 & 1.23 & ~ & 4.567 & 20.70 & 1.73  \\ \hline
        2.700 & -1.16 & 0.46 & ~ & 3.333 & -3.24 & -0.49 & ~ & 3.967 & 12.10 & 1.78 & ~ & 4.600 & 20.30 & 2.00  \\ \hline
        2.733 & -1.02 & -0.03 & ~ & 3.367 & -3.18 & 0.08 & ~ & 4.000 & 12.71 & 2.12 & ~ & 4.633 & 20.18 & 1.11  \\ \hline
        2.767 & -0.45 & 0.09 & ~ & 3.400 & -2.91 & -0.09 & ~ & 4.033 & 13.54 & 1.42 & ~ & 4.667 & 20.85 & 1.14  \\ \hline
        2.800 & -1.14 & -0.05 & ~ & 3.433 & -1.40 & 0.15 & ~ & 4.067 & 13.82 & 2.04 & ~ & 4.700 & 21.00 & 0.98  \\ \hline
        2.833 & -1.11 & 0.11 & ~ & 3.467 & -0.10 & 0.93 & ~ & 4.100 & 14.64 & 1.42 & ~ & 4.733 & 21.22 & 1.53  \\ \hline
        2.867 & -2.24 & 0.19 & ~ & 3.500 & 0.80 & 0.27 & ~ & 4.133 & 15.54 & 1.65 & ~ & 4.767 & 21.38 & 1.29  \\ \hline
        2.900 & -2.53 & -0.48 & ~ & 3.533 & 0.95 & 1.09 & ~ & 4.167 & 16.84 & 1.48 & ~ & 4.800 & 20.88 & 0.40  \\ \hline
        2.933 & -2.96 & -0.05 & ~ & 3.567 & 0.75 & 1.23 & ~ & 4.200 & 17.00 & 2.03 & ~ & 4.833 & 21.17 & 1.85  \\ \hline
        2.967 & -2.37 & -0.04 & ~ & 3.600 & 1.15 & 1.01 & ~ & 4.233 & 17.33 & 2.07 & ~ & 4.867 & 21.60 & 1.90  \\ \hline
        3.000 & -2.75 & 0.28 & ~ & 3.633 & 2.77 & 1.55 & ~ & 4.267 & 18.09 & 1.25 & ~ & 4.900 & 21.72 & 1.35  \\ \hline
        3.033 & -2.94 & -0.04 & ~ & 3.667 & 3.19 & 0.16 & ~ & 4.300 & 18.93 & 2.05 & ~ & 4.933 & 21.35 & 2.04  \\ \hline
        3.067 & -3.51 & 0.71 & ~ & 3.700 & 4.96 & 1.59 & ~ & 4.333 & 19.11 & 1.26 & ~ & 4.967 & 20.64 & 2.15  \\ \hline
        3.100 & -4.36 & -0.08 & ~ & 3.733 & 6.04 & 1.80 & ~ & 4.367 & 19.44 & 0.96 & ~ & 5.000 & 21.07 & 1.88  \\ \hline
        3.133 & -4.17 & 0.00 & ~ & 3.767 & 5.40 & 1.75 & ~ & 4.400 & 19.53 & 0.99 & ~ & 5.033 & 21.66 & 2.37  \\ \hline
    \end{tabular}
    \caption*{Continued}
\end{table}

\begin{table}[!ht]
    \centering
    \begin{tabular}{|l l l|l|l l l|l|l l l|l|l l l|l|l l l|}
    \hline
        time & x & y & ~ & time & x & y & ~ & time & x & y  \\ \hline
        5.067 & 21.81 & 1.45 & ~ & 5.700 & 20.74 & 0.62 & ~ & 6.333 & 21.21 & 2.93  \\ \hline
        5.100 & 21.78 & 1.83 & ~ & 5.733 & 20.41 & 2.14 & ~ & 6.367 & 22.02 & 2.69  \\ \hline
        5.133 & 22.15 & 2.28 & ~ & 5.767 & 20.55 & 1.59 & ~ & 6.400 & 21.94 & 2.44  \\ \hline
        5.167 & 22.32 & 1.87 & ~ & 5.800 & 20.01 & 1.83 & ~ & 6.433 & 21.28 & 2.61  \\ \hline
        5.200 & 21.55 & 1.79 & ~ & 5.833 & 20.35 & 2.29 & ~ & 6.467 & 21.64 & 2.37  \\ \hline
        5.233 & 21.22 & 1.64 & ~ & 5.867 & 21.03 & 3.17 & ~ & 6.500 & 21.39 & 2.11  \\ \hline
        5.267 & 21.15 & 1.63 & ~ & 5.900 & 20.70 & 2.35 & ~ & 6.533 & 21.19 & 2.31  \\ \hline
        5.300 & 21.07 & 1.56 & ~ & 5.933 & 21.09 & 1.81 & ~ & 6.567 & 21.22 & 1.78  \\ \hline
        5.333 & 21.07 & 0.95 & ~ & 5.967 & 20.67 & 2.05 & ~ & 6.600 & 21.89 & 1.83  \\ \hline
        5.367 & 20.40 & 1.56 & ~ & 6.000 & 20.55 & 1.84 & ~ & 6.633 & 21.14 & 2.15  \\ \hline
        5.400 & 20.56 & 1.16 & ~ & 6.033 & 20.65 & 2.11 & ~ & ~ & ~ &   \\ \hline
        5.433 & 20.60 & 0.95 & ~ & 6.067 & 20.46 & 2.53 & ~ & ~ & ~ &   \\ \hline
        5.467 & 20.91 & 2.06 & ~ & 6.100 & 21.34 & 2.03 & ~ & ~ & ~ &   \\ \hline
        5.500 & 21.13 & 2.40 & ~ & 6.133 & 21.43 & 2.46 & ~ & ~ & ~ &   \\ \hline
        5.533 & 20.92 & 1.99 & ~ & 6.167 & 21.45 & 2.08 & ~ & ~ & ~ &   \\ \hline
        5.567 & 21.09 & 2.05 & ~ & 6.200 & 21.11 & 2.16 & ~ & ~ & ~ &   \\ \hline
        5.600 & 20.27 & 1.58 & ~ & 6.233 & 20.65 & 1.63 & ~ & ~ & ~ &   \\ \hline
        5.633 & 20.94 & 1.66 & ~ & 6.267 & 20.09 & 2.40 & ~ & ~ & ~ &   \\ \hline
        5.667 & 20.63 & 1.89 & ~ & 6.300 & 20.96 & 2.92 & ~ & ~ & ~ &   \\ \hline
    \end{tabular}
    \caption*{Continued}
\end{table}
   
\newpage

\printbibliography
\end{document}